# A Hybrid Usability Approach for Rating Evaluation of M-Commerce Applications

Ahmad Ibtisam
Department of Computer Science
*University of Lahore*
Lahore, 55150, Pakistan
Ibti438@gmail.com

Muhammad Bilal Khan
Faculty of Information Technology
*University of Central Punjab*
Lahore, Pakistan

Arshad Ali
FAST School of Computing
National University of Computer and Emerging Sciences, Pakistan
arshad.ali1@.edu.pk

nu

***Abstract*— The success of any mobile application relies on its usefulness and rating is considered as an important measure in this regard. This research work focuses on identifying usability factors, which contribute significantly towards the rating of M-commerce apps. This work intends to explore existing usability models consisting of different factors along with a set of criteria and evaluate in terms of rating estimation by considering 5 well-known mobile applications, namely (i) daraz, (ii) shophive, (iii) home shopping, (iv) Symbios. (v) yayvo. Then, this work provides a hybrid usability model for rating prediction of M-commerce applications. The initial hybrid usability model comprises of (i) learnability, (ii) consistency, (iii) human factors,(iv)communicativeness,(v)effectiveness, (vi) Operability, (vii) efficiency, (viii) satisfaction. Each factor consists of some criteria. Keeping in view the factors of hybrid usability model, the data was collected from 40 users for each application. Furthermore, Forward Stepwise Multiple Linear Regression based rating prediction model is suggested by analyzing each criterion of all factors of hybrid usability model. Finally, the model is assessed and validated by using PRED(x) and K-fold techniques.**

## I. INTRODUCTION

Digital technology or consumer-based shopping market, electronic commerce has evolved over the years and has profoundly affected individuals and organizations. Generally, E-commerce consists of maintaining relationships and carrying out commercial transactions that include the sale of information, services and goods through computer telecommunications networks. E-commerce originated in a standard for the exchange of commercial documents, such as orders or invoices, between suppliers and their commercial customers [1]. Electronic commerce consists of commerce between companies, consumers and in internal organizational transactions that support such activities [1]. Electronic markets are places in cyberspace where potential sellers and buyers can carry out transactions in an effective way through interactions in electronic networks [2]. Transactions between companies and consumers involve exchanges of goods and services that are carried out through computer networks. Companies like Alibaba [1] have created business-to-business markets. It involves commercial companies on both sides of the transactions, for example, when a company sends an order from a supplier [2]. The consumer-to-consumer markets include large electronic shopping centers such as Amazon and auction platforms such as eBay. The transactions are carried out between individuals [2]. The e-commerce technology has made tremendous progress during the past few years. It has become part of everyday life for most people. The consumers no longer need to visit stores or market places physically, in order to buy some products; which can now be done with a small device in the palm of the hand known as a mobile phone while sitting at home. Mobile commerce is the fastest growing form of e-commerce and the main areas of growth are retail sales, with an expansion at a rate of 50% or more per year [3]. M-commerce is about the advancement of applications and services that have become accessible from mobile devices enabled for the Internet. It's about new technologies, services and business models and it somewhat varies from traditional e-commerce. However, it is not always easy to define M-commerce [14]. What is important when defining mobile commerce is that at least the initiation and completion of the transaction is done through mobile access through a mobile device. Gale [4] points out that M-commerce consists of some unique characteristics with certain advantages over conventional forms of commercial transactions. These advantages include real-time data and information access, authenticity, ubiquitous features, personalized contents, numerous communication functionalities and constant access. Since mobile commerce makes routine tasks more convenient, the use of mobile phones continues to grow. As per published report [16], up to 65% of the Swedish population is connected to their smartphone daily. According to [17], for a brief period, the smartphone has fundamentally changed consumer behavior, both in electronic commerce and in physical commerce. However, it is continuously moving toward M-commerce. Nowadays everyone is dependent on wireless handheld devices and M-commerce is mainly based on wireless devices like smartphones. Moreover, to ensure the quality of these applications there are some software quality models. Software Quality Models (SQM) recognized as a standard method for measuring the quality of a software product [6]. Research community has focused a lot and proposed various SQMs [7-12], [13], [18-20] with respect to product operation. McCall's model [10], states that the product's operating category consists of five software quality factors, which deal with the requirements that directly affect the day-to-day operation of the software. These are correctness, reliability, efficiency, integrity, usability. Usability is observed, as the most important factor of various SQMs. Usability is the ease of use and learning ability of man-made objects, such as tools or devices. When considering the ease of use in the application, the application can be interested in a way that meets the needs of the consumer and makes the application easier and better, which help in sales growth [6]. However, mobile phones can provide access to many new applications, but they also have limitations such as small screen size, limited screen resolution and the most difficult input mechanisms. Surveys of mobile Internet users show that ease of use is the biggest source of frustration [15]. It can be an important step for companies to adopt M-commerce, which may be even more important for them to develop mobile applications for M-commerce correctly. There are many factors to be considered when developing mobile applications or sites that support mobile commerce, and ease of use is one of the most critical. In fact, Mobile phones brought new challenges by creating a good user experience and ease of use. This work first provides a hybrid usability model and then presents a regression-based model for rating prediction by analyzing each criterion of all the factors of hybrid usability model. This study also made an effort to identify issues which can be addressed during the design phase of M-commerce applications.

### *A. PROBLEM DESCRIPTION*

During the investigation of mobile commerce, we found that the trend of smartphones continues to increase, and that Pakistan is one of the leading countries in the use of the mobile league. However, the rating prediction of M-commerce application rating has not been given that much attention by the researchers. Similarly, a systematic approach for prediction of mobile application rating is essential for improving application design. Therefore, we feel that the factors that determine the rating are more important as compared to the rating itself.

### *B. OBJECTIVES*

The purpose of this work is to fill the research gap by investigating the impact of various usability factors for determining the rating of M-commerce applications.
The main objectives of this work are as follows:

- ➔ *To explore various usability models and their factors.*
- ➔ *To provide a hybrid usability model for determination of rating of various M-commerce applications.*
- ➔ *To develop a model meant for rating prediction of android-based M-commerce applications based on various usability factors through regression analysis.*

### *C. RESEARCH METHODOLOGY*

Research is basically described as a systematic and gradual description of the problem under consideration through the chosen method for the purpose of finding their solution. Firstly, various existing usability models were studied and investigated in order to find a research problem. Furthermore, this work focuses on a quantitative approach for collecting primary data through the distribution of a well-structured survey questionnaire, which was derived from various existing instruments developed by well-known researchers.

All items of the questionnaire are arranged in the form of a 5-point Likert scale (1-strongly disagree, 5 = strongly agree) in order to make them understandable and consistent. The population of the study comprise customers using M-commerce applications for the purpose of purchasing various products in Pakistan. Thus, we have 200 users for five applications. However, in analysis we considered 168 instances. Initially, there were 144 questions, and some were reduced to 60 questions of various usability factors by using a Delphi approach [21].

The other step was to identify the attributes and analyze the identified attributes by applying linear and multiple regression for rating estimation. Lastly, the validation and assessment of the model was carried out.

The structure of the remaining paper is as follows: Section II provides literature review with emphasis on related work. Section III discusses the Delphi process and then elaborates the data collection part. Section IV presents a prediction model. Section V is about model validation and assessment. Finally, Section VI concludes the work.

## II. LITERATURE REVIEW

In the past 30 years, a series of software quality models have been developed. Quality models represent a quantitative structure of quality. This section presents the related literature of quality models with usability as a prime attribute. The first part elaborates the software quality models. The second part contains information on quality models developed for mobile applications.
Jim McCall presented a software quality model [10], which is one of the most famous predecessors of current quality models. McCall developed the model for the US Army, UU in 1977. With respect to product operation, McCall's quality model consists of the following factors: Correction, Efficiency, Integrity, Reliability and Usability.
Robert Grady and Hewlett Packard presented the FURPS model [10]. They built the FURPS model for Rational Software Company. FURPS model comprises the following non-functional requirements: Usability, Reliability, Performance, and Compatibility.
Furthermore, researchers also considered ISO standards i.e. ISO 9126, ISO 9142, ISO 25000, ISO 25010 with usability as a major factor. However, Quality model Bertoa [13] depends on the ISO 9126 Model. It characterizes a lot of quality properties for the effective evaluation of COTS. COTS are utilized by software development organizations to make increasingly complex software.
According to the Bertoa, usability comprises of various factors namely learnability, understandability, and Operability. Learnability is the mechanism by which a system can recognize what sort of thing you are afterwards. Understandability means how users can choose a software product that is appropriate for their use and how it can be utilized for specific tasks. Operability is the ability of the software product to allow the user to operate and control it.
Kurosu proposed his latest model as ISO/IEC25010 [19]. According to Kurosu-3, usability comprises of recognizability, learnability memorability, user error protection, operability, and accessibility.
Recognizability of a product or system can be used by specific users for achieving specific goals with effectiveness, efficiency, and satisfaction in a specified context of use. Learnability of a product or system empowers the client to figure out about its utilization with effectiveness, efficiency and in critical circumstances. Memorability is the degree to which users can recall how to use the provided interface. User error protection is also the degree in which a product or system is anything but difficult to work, control and fitting to utilize. Operability is the ability of the software product to allow the user to operate and control it. Accessibility is the degree in which a product or system can be utilized by individuals with the broadest scope of qualities and abilities to accomplish a predetermined objective in a predefined set of utilization.

### *A. RELATED WORK*

Hartmut Hoehle (at el) [25] researched on the usability of social media applications. In this paper, different cultural qualities aspects were examined; manliness independence/, power separation, vulnerability evasion, and long haul direction [25]. These cultural attributes are utilized as moderators to test a model to which extent the effect of social media application of usability that fulfils needs of people with different aspects. The model-included information gathered from 1844 customers from 4 nations. The conclusions presented by this model clarified 38% of the change in it aimed to utilize them in terms of social media apps. This article provides multidimensional research by looking at all five mentioned national cultural qualities as possible factors concerning mobile social media applications.

Ching-Wen Hsu et al [26] make research which is going to help to work on basic elements of usability for the effectiveness of M-Commerce selection. In this paper, Mobile trade have consistently been analyzed on technology acceptance model (TAM) which is accomplished by using multiple conditions and modelling. The reason for this research was to discover basic influencing factors toward trade carried out by Mobile Commerce, which would impact the customer's choice on M-Commerce platforms. This analysis utilizes the DEMATEL strategy to recognize those elements that are affecting to which extent. DEMATEL is a 5-stage process, to be specific, generating, normalizing, and achieving the direct impact framework method. The critical success factors that are found to be influencing the model more were perceived ease of using them, perceived usefulness of Mobile Commerce, value-added services, and service functionality. The consequences of this investigation are going to explain that M-trade specialist is encouraging advancement in the field of Mobile Commerce trade. Next, the assessment elements are utilized in the relevant writing on fractional TAM factors, which was going to exclude all those factors, which were giving negative correlations toward M-business.

Hartmut Hoehle et al [27] introduced his research paper on mobile usability factors and policies that guide toward their usage. This study conceptualizes mobile application usability creates and approves an instrument to validate their measurement to which level it is measured to be the best. This explains that Mobile application usability has been gaining a lot of attention of the people across the world when we consider the human's interactions with PC [27]. To conceptualize mobile application usability, Hartmut researched Microsoft's mobile usability rules that they defined in the latest mobile application development in the market and emphasize on the better enhancement from the perspective of usability. After that the author has gone through the pilot study of the multiple magazines about the applications and their usability and used SPSS technique in putting all content gathered from multiple magazines by this pilot study. When doing statistical analysis on exploratory factor investigation techniques that apply to the German people. Therefore, that influences them to purchase and utilize mobile social media applications on their cell phones. The results showed that mobile application usability is predicting the behaviors of German people this paper can be utilized to direct future research in human to PC interaction and will help in the feasible structure of mobile applications. Kim et al [28] present a paper, which is going to elaborate on the relation of usability factor and product success factor in mobile applications. The objective of this research study was to create a questionnaire for mobile phones and find out what could be the relation between usability and the success of the product in the market. Increase in purchasing intentions of the people and success scores is also including usability factor that was analyzed to determine if they are carrying positively correlation with each other or not. With multiple linear regression analysis MLR, six main factors have been evaluated as that may affect cell phone products going to be playing their role in success. They are designed in such a way that meets the customer's requirements, satisfaction, latest techniques which were newly introduced, feedback of the customer is integrated into that as well as the efficiency of the application. The survey was conducted among a total of 219 candidates and involved 13 usability factors and 8 success factors. The only limitation of this study is that it was not brand or model-oriented.

D. Lee et al [29] worked on finding the relationship between usability with simplicity factors. In this paper, research was conducted on how humans interact with mobile phones and how to gain access to different functions in the easy way [29]. 310 mobile clients from Korea participated in this research. The technique that was used to conduct research was utilizing Partial Least Squares (PSL) technique, as it is more reasonable than the covariance-based methodology. Simplicity factor and interaction factor were taken to be two variables defined in this research that need to be analyzed in the research. The reliability factor for both of the variables was higher than 0.80 and the AVE was higher than 0.50, which is predicting the strong reliability and validity of the data as well as in variables. Karima Moumane et al [30] processed an experimental analysis dependent on a lot of measures to assess the usability of mobile applications running on various mobile operating systems and see if one is the better to influence the customer. In this analysis usability of mobile applications is assessed on the multiple operating systems; IOS, Android, Symbian. A model is created on software quality standards of the ISO 9126 which is utilized to assess which one of the operating systems is better. ISO 25062 and ISO 9241 models have been utilized in this research and consist of information collected from 32 clients, which 7.0 questionnaires. Two applications were utilized in this investigation; google applications and google maps. Multiple difficulties were recognized while utilizing applications by the users and the flaws have been identified one of them was screen size, the resolution of the screen and the memory of the mobile phone, which approve the discoveries of our system. There are also some software issues that have been identified which were the presence of online help and user guide were not found to be compatible with the user.

Ali Balapour et al [31] published a research paper Meta-regression study on the usability of applications. This article was going to investigate the correlation between usability perception and a factor influencing usability. The information provided for usability and application design both impact the usability of the user. Conversely, application design has been identified as a major issue in the usability of applications but no observations was limited too.

## III. DELPHI PROCESS

To understand the behavior of M-commerce applications, we adopted a questionnaire used previously in different studies developed by different researchers [3]. The usability questionnaire consists of 60 items with 5-point Likert scale on 13 usability attributes namely efficiency, learnability, satisfaction, human factors, communicativeness, effectiveness, operability, protection, consistency, accessibility, user errors protection, memorability recognisability. The usability questionnaire was evaluated in two rounds by using a Delphi approach [21]. For this reason, eight experts participated. To receive good results experts gave comments on all items, scale, purpose which resulted in correction of some parts. After adding, eliminating and refining, Appendix A presents the final questionnaire. The Delphi approach is commonly used for gathering data from surveys [22]. The reason behind the Delphi approach is to have more opinions. That means, a group of experts can provide a better response as compared to individual result [21 ].

According to the Vidal [22], participants should be 8-18 to accomplish the significant results and to evade difficulty. During the selection of the experts keep in mind that experts should have enough experience, willingness, sufficient time and good communication skills to participate in Delphi [23]. For this purpose, the qualification of the experts matters a lot. The panel experts consist of lecturers and Assistant professors having appropriate knowledge of usability. For this study, a two-round Delphi survey was performed, as the group, the consensus is required. Purpose of the research depends on the number of rounds to perform this approach [24].

In the first and second round all the experts were the same. In the first round of the Delphi, approach data was gathered in two sections; demographics were asked through questionnaire and were validated in a 5-point scale and second section based on the attributes of usability. That helps to determine the role of each attribute on M-commerce and those questionnaires were also asked by using a 5-point scale. Because of the first round, demographics were acquired from the experts. After those 144 general questionnaires asked on usability in M-commerce from the expert on 5- point scale. Each point contains 20% weighted. All the frequencies were calculated. On the scale 4 and 5 (agree and strongly agree) considered as in the support of questions that were asked. All the experts gave their comments and suggestions on the questionnaire. The suggestions were:

Rephrase the questions and some comments on grammatical errors. Changing neutral to don't know in Likert scale to be clearer Reduce questions because it is complicated to fill too many questions.

In the 2nd round after removing and improving items according to the expert's opinions. Thus, the factors are lower than 70%, a mean rate lower than 3.2 were deleted from the questionnaire. Afterwards, 60 items were left that fulfils the criteria.

### *A. DATA COLLECTION*

After performing the Delphi approach, we have finalized 40 questions of various usability factors. A finalized questionnaire was used in this study, in which the chosen participants took part in assessment M-commerce mobile app by conveying their thoughts while carrying out set tasks. We made sure that all participants execute the tasks in order to accurately evaluate its functionalities and design. The reason for selecting this method is it being less expensive and providing results that are close to what is experienced by users [24].

The questionnaire was solved by using two methods i.e. web-based questionnaire and by providing a controlled environment to the group of participants. In a web-based questionnaire, a questionnaire was designed on a web site using google forms. A participant received an email which contains a link. The user was directed toward the web page for his/her response. However, in the control environment, the number of users that can be undergraduate, graduate or lecturer were gathered in a room. That was a proper sitting environment arrangement. Each participant had an android phone with an M-commerce application. In printed form, a questionnaire was provided, and they performed the different tasks according to the questionnaire or instructed by us. All the questionnaires were collected from the participants to analyze the attribute. The total number of 200 instances were gathered from participants. 32 were rejected, as they were not appropriate and not properly filled. Thus, the remaining 168 instances were used for analysis purposes.

## IV. PREDICTION MODEL

Researchers are going to use the approach and procedures adopted by the different people for the analysis of the predictors rating according to usability. Predictions are made on the base of the regression model, and then comparison made with the hybrid model for rating. K-Fold, MRE inclusive of the MMRE technique to assess and validate the proposed model.

In modeling techniques, there are three different MLR techniques. These techniques are forward selection, backward selection and stepwise. Researchers used Stepwise regression as our modeling technique to build our prediction model. We have applied stepwise over the data of sample size 168. It is the most commonly used technique in literature due to its performance level. It is used to depict the relationship between dependent variable with the independent variables.

### *A. SLR ANALYSIS*

Initially, Simple linear regression was used to analyze the individual strength of possible eight predictors. After that Multi linear regressions (MLR) technique was used. However, this model is used for rating as well as the determination of usability factors that tells which one is significantly effective [32]. Regression analysis was performed with the help of SPSS. Table 1 is presenting the results of SLR. Furthermore, these results show that in mobile applications efficiency is the most reliable predictor whereas the human factor considered as the weakest predictor in all the usability factors. We examined that efficiency is the most reliable predictor of usability factors.in mobile applications Here we define coding for variables namely effectiveness as EFFI, learnability as LERN, communicativeness as COMM, human factor as HUFA, consistency as CONS, operability as OPER, satisfaction as SATS and efficiency as EFFI.

### *B. MLR ANALYSIS*

The set of data containing different instances (8 individual usability factors) which were gathered for model prediction. Stepwise multilinear regression were performed by putting all multiple independent variables with the dependent variable i.e., two, three up to 8 factors as independent variables. We've analyzed the values of $R^2$ and their significance. The result of these predictors shows significant value 0.05 by using MLR. Furthermore, Variance Inflation Factor was used to check the multi collinearity between usability predictors.

Stepwise regression for the combinations of 7 variables is shown below in Table 1. MLR model with seven predictors is showing the greater value of the R2 that is 0.910. This model shows 91% variation in determination of user rating prediction for usability factors. Therefore, the final suggested model turns out to be with 7 factors, namely (i) Effectiveness, (ii) Learnability, (iii) Communicativeness, (iv) Consistency, (v) Operability, (vi) Satisfaction, (vii) Efficiency. The regression equation of the suggested model is given by:

$$\hat{Y} = \mathbf{0.078 + (0.136 * EFFE) + (0.340 + LERN) + (-0.088 + COMM) + (-0.160 + CONS) + (0.182 * OPER) + (0.105 * SATS) + (0.333 * EFFI)}$$

Here, $\hat{Y}$ represents model predicted value of rating.

Table 1. MLR with 7 predictors

| **Predictors** | **$R^2$** | **Unstandarized Beta( Constant)** | **Standarized Beta** | **Sig.** | **VIF** |
|---|---|---|---|---|---|
| EFFI | 0.910 | 0.078 | 0.333 | 0.000 | 3.261 |
| LERN | | | 0.340 | 0.000 | 2.345 |
| EFFE | | | 0.136 | 0.002 | 3.428 |
| CONS | | | -0.160 | 0.000 | 2.126 |
| OPER | | | 0.182 | 0.000 | 1.363 |
| SATS | | | 0.105 | 0.000 | 1.375 |
| COMM | | | -0.088 | 0.006 | 1.762 |

In MLR analysis, we gathered all the results of possible sets containing two, three, four, five, six, seven and eight predictor variables. We added the variables gradually until we got optimal model. Table 2 shows the complete summary of most significant models for each group of combinations.

Table 2: Summary of MLR Based Models

| **Sr.** | **Predictors** | **Prediction Model** | **$R^2$** |
|---|---|---|---|
| **No** | | | |
| 1 | EFFI | $\hat{Y} =$ 0.347 + (0.846*EFFI) | O.715 |
| 2 | EFFI, LERN | $\hat{Y} =$ -0.248+ (0.487* LERN) + (0.517*EFFI) | 0.844 |
| 3 | LERN, CONS and EFFI | $\hat{Y} =$ 0.498+ (0.452*LERN) + (-0.182*CONSIS) + (0.441*EFFI) | 0.867 |
| 4 | LERN, CONS, EFFI and OPER | $\hat{Y} =$ 0.344+ (0.376*LERN) + (-0.242* CONS) + (0.165* OPER) + (0.452* EFFI) | 0.889 |
| 5 | EFFE, LERN, CONS, OPER and EFFI | $\hat{Y} =$ 0.113 + (0.170*EFFE) + (0.366 * LERN) + (-0.193*CONS) + (0.152*OPER) + (0.352*EFFI) | 0.898 |
| 6 | EFFE, LERN, CONS, OPER, SATS and EFFI | $\hat{Y} =$ -0.24+ (0.168* EFFE) + (0.331* LERN) + (-0.192* CONS) + (0.158* OPER) + (0.101* SATS) + (0.332* EFFI) | 0.906 |
| **7** | **EFFE, LERN, COMM, CONS, OPER, SATS and EFFI** | $\hat{Y} =$ **0.078+ (0.136* EFFE) + (0.340* LERN) + (-0.088* COMM) + (-0.160* CONS) + (0.182* OPER) + (0.105* SATS) + (0.333* EFFI)** | **0.910** |

### *C. HYBRID VS SUGGESTED MODEL*

A hybride purpose of knowing the improvement in terms of rating prediction as a result of suggested regression model, a hybrid model comprising of all 8 factors was evaluated. The average of all 8 factors for each instance was taken as a hybrid model. Our suggested MLR model for rating prediction consists of all factors just like hybrid model except one factor i.e., human factor (HUFA).

Our suggested model performs better in terms of rating prediction, as the same is evident from the value of the coefficient of determination ($R^2$). It has been observed that 91% of the variation in user rating prediction is based on the seven usability factors ($R^2$= 0.910) as compared to Hybrid model which explains only 0.335% variation ($R^2$ = 0.335).

## V. MODEL ASSESSMENT AND VALIDATION

Infact, normal metrics of prediction were applied to reach the remarkable accuracy of the predicted model. For this purpose, MMRE and PRED(x) were used to find out the accuracy of our model. Moreover, in literature values of x commonly used as 25. Meanwhile, PRED (25) must be equal to or more than 0.75 and MMRE must be less than and equal to 0.25 for the validation of suggested model for accuracy [33]. MMRE turns out to be 0.0515.

Where Y is the actual value (user rating) and $\hat{Y}$ is predicted (model value).

### *A. K-FOLD CROSS VALIDATION*

All folds comprise the data values divided into 8 fixed groups. By performing each fold, every group consists of 21 instances as training data i.e., 1-21, 22-42, 43-63, 64-84, 85-105, 106-126, 127-147 and 148-168 for folds from K = 1 to

K =8 respectively. All the data values other than those mentioned before were used to train the model during each iteration. By doing this, we noticed that values were significant which are equals to or less than 0.05. So, all of the seven predictors proved significant values. Moreover, the PRED (25) values of all folds were equivalent to one. Table 3 depicts the equations and $R^2$ of all the fixed folds and shows that the suggested model is validated quite well.
For this purpose, fixed folds were made which shows the accuracy and significance of the result. Table 3 depicts the equations and $R^2$ of all the fixed folds. It has been observed that all the K-FOLDs for all instances i.e., 1-21, 22-42 and so on are showing significant results with good $R^2$.

Table 3: Summary of fixed K-folds

| Fold No. | Prediction Models | $R^2$ |
|---|---|---|
| 1. | Ŷ (K-1) = .121 + (0.119*EFFE) + (0.350*LERN) + (-0.107*COMM) + (-0.157*CONS) + (0.186*OPER) + (0.099*SATS) + (0.339*EFFI) | 0.914 |
| 2. | Ŷ (K-2) = .079 + (0.100*EFFE) + (0.388*LERN) + (-0.084*COMM) + (-0.160*CONS) + (0.155 * OPER) + (0.132*SATS) + (0.332*EFFI) | 0.915 |
| 3. | Ŷ (K-3) = .212 + (0.097*EFFE) + (0.320*LERN) + (-0.094*COMM) + (-0.195*CONS) + (0.197*OPER) + (0.108*SATS) + (0.374*EFFI) | 0.908 |
| 4. | Ŷ (K-4) = .050 + (0.123*EFFE) + (0.360*LERN) + (-0.089*COMM) + (-0.160*CONS) + (0.174*OPER) + (0.101*SATS) + (0.336*EFFI) | 0.905 |
| 5. | Ŷ (K-5) = .089 + (0.141*EFFE) + (0.327*LERN) + (-0.091*COMM) + (-0.160*CONS) + (0.193*OPER) + (0.105*SATS) + (0.321*EFFI) | 0.911 |
| 6. | Ŷ (K-6) = -.001 + (0.155*EFFE) + (0.338*LERN) + (-0.070*COMM) + (-0.146*CONS) + (0.173*OPER) + (0.088*SATS) + (0.326*EFFI) | 0.914 |
| 7. | Ŷ (K-7) = -.004 + (0.160*EFFE) + (0.322*LERN) + (-0.084*COMM) + (-0.140*CONS) + (0.178*OPER) + (0.109*SATS) + (0.326*EFFI) | 0.909 |
| 8. | Ŷ (K-8) = .081 + (0.196*EFFE) + (0.311*LERN) + (-0.079*COMM) + (-0.165*CONS) + (0.189*OPER) + (0.103*SATS) + (0.301*EFFI) | 0.910 |

After that, eight folds of the randomly arranged instances was done. Therefore, the data collected in the SPSS to check its significance and $R^2$. So, even in random k-fold all the folds are found to be significant and giving good values of $R^2$ which is predicting a fact that all the data is valid even when abruptly arranged in SPSS. Table 4 depicts the equations and $R^2$ of all the random folds.

Table 4: Summary of random K-Fold

| Fold No. | Prediction Models | $R^2$ |
|---|---|---|
| 1. | Ŷ (K-1) = .105 + (0.103*EFFE) + (0.353*LERN) + (-0.072 * COMM) + (-0.175*CONS) + (0.175*OPER) + (0.119 * SATS) + (0.338*EFFI) | 0.908 |
| 2. | Ŷ (K-2) = .091 + (0.153*EFFE) + (0.341*LERN) + (-0.091 * COMM) + (-0.155*CONS) + (0.184 * OPER) + (0.094 * SATS) + (0.318*EFFI) | 0.914 |
| 3. | Ŷ (K-3) = .059 + (0.145*EFFE) + (0.344*LERN) + (-0.083 * COMM) + (-0.180*CONS) + (0.203 * OPER) + (0.117 * SATS) + (0.297*EFFI) | 0.914 |
| 4. | Ŷ (K-4) = .016 + (0.145*EFFE) + (0.353*LERN) + (-0.080 * COMM) + (-0.126*CONS) + (0.154 * OPER) + (0.079 * SATS) + (0.356*EFFI) | 0.909 |
| 5. | Ŷ (K-5) = .121 + (0.140*EFFE) + (0.305*LERN) + (-0.085 * COMM) + (-0.172*CONS) + (0.190 * OPER) + (0.093 * SATS) + (0.363*EFFI) | 0.907 |
| 6. | Ŷ (K-6) = .077 + (0.143*EFFE) + (0.325*LERN) + (-0.103 * COMM) + (-0.163*CONS) + (0.195 * OPER) + (0.108 * SATS) + (0.338*EFFI) | 0.903 |
| 7. | Ŷ (K-1) = .093 + (0.133*EFFE) + (0.335*LERN) + (-0.090 * COMM) + (-0.172*CONS) + (0.179 * OPER) + (0.115 * SATS) + (0.332*EFFI) | 0.915 |
| 8. | Ŷ (K-1) = .060 + (0.150*EFFE) + (0.343*LERN) + (-0.096 * COMM) + (-0.137*CONS) + (0.172 * OPER) + (0.098 * SATS) + (0.328*EFFI) | 0.910 |

## VI. CONCLUSION

Mobile Commerce applications have been evaluated based on various possible aspects of usability. As most of the people are using mobile applications in their daily life for shopping and many other useful activities. For this purpose, data was collected from users of M-commerce apps in order to However, rating is considered as the important part to evaluate any applications. A hybrid usability factor-based model was derived and applied to judge which factor has more impact on rating. During the first phase, factors were identified. The factors that need to be evaluated for determination of rating are (i) Efficiency (ii) Learnability (iii) Satisfaction (iv) Human factor (v) communicativeness (vi) effectiveness (vii) operability (viii) consistency. Each factor consists of some criteria.
Then, SLR was applied on the basis of 8 factors i.e., (i) EFFI (ii) LERN (iii) SATS (iv) HUFA (v) COMM (vi) EFFE (vii) OPER (viii) CONS. It was observed that all factors provided significant results with good $R^2$ values. It was observed that EFFE, LERN, CONS and EFFI are the usability factors having a great impact on rating determination.
Next, stepwise regression was performed. The combination of learnability and efficiency showed maximum $R^2$ as 0.844 i.e., it explains 84.4% of the variation in the model. Moreover, the best model turned out to be with seven predictors i.e., EFFE, LERN, COMM, CONS, OPER, SATS and EFFI with $R^2$ as 0.910. The suggested regression-based model with 7 variables was compared with hybrid model comprising of 8 variables and it was noticed that regression-based model outperformed hybrid model.
For Model Assessment, MMRE and PRED(x) were used while K-Fold cross validation was used for model validation. The model was assessed well and then it was observed that results of all folds (fixed as well as random) were significant with $R^2$ up to or more than 0.9. For each, MMREs results are satisfactory as well. In this way, our model was validated.
determine the rating of these applications. Results were obtained with respect to the usability of M-commerce applications.

**References**

1. Gunasekaran, A., Marri, H. B., McGaughey, R. E., & Nebhwani, M. D. (2002). E-commerce and its impact on operations management. *International journal of production economics*, *75*(1-2), 185-197.
2. Choshin, M., & Ghaffari, A. (2017). An investigation of the impact of effective factors on the success of e-commerce

in small and medium-sized companies. *Computers in Human Behavior*, *66*, 67-74.
3. Licina, A., Radtke, H., & Johansson, C. (2018). Usability in M-commerce: Critical factors to consider when adopting m-commerce.
4. Gale. (2009). Mobile Commerce. Encyclopedia of Management, 6 .ed., pp. 582-585.
5. Venkatesh, V., Ramesh, V., & Massey, A. P. (2003). Understanding usability in mobile commerce. *Communications of the ACM*, *46*(12), 53-56.
6. Hussain, A. O., & Mkpojiogu, E. M. (2016). Usability evaluation techniques in mobile commerce applications: A systematic review. Proceedings of the International Conference. Applied Science and Technology, 1761 (1).
7. Nielsen, J. (1994). Usability engineering. California: Elsevier Science.
8. Zhang, D., & Adipat, B. (2005). Challenges, methodologies, and issues in the usability testing of mobile applications. International Journal of Human-Computer.
9. Malhotra, N. & Pruthi, S. (2012). Efficient Software Quality Models for Safety and Resilience. International Journal of Recent Technology and Engineering (IJRTE). 1 (3). 66-70
10. Suman. & Wadhwa, M. (2014). A Comparative Study of Software Quality Models. International Journal of Computer Science and Information Technologies (IJCSIT). 5 (4).
11. Abran, A., Khelifi, A., Suryn, W., & Seffah, A. (2003, April). Consolidating the ISO usability models. In Proceedings of 11th international software quality management conference (Vol. 2003, pp. 23-25).
12. Measuring public value UX-based on ISO/IEC 25010 quality attributes: a Case study on e-Government website. In User Science and Engineering (i-USEr), 2014 3rd International Conference on(pp. 56-61). IEEE.
13. Spiekermann, S., Grossklags, J., & Berendt, B. (2001, October). E-privacy in 2nd generation E-commerce: privacy preferences versus actual behaviour. In Proceedings of the 3rd ACM Conference on Electronic Commerce (pp. 38-47). ACM.
14. Tarasewich, P., Nickerson, R. C., & Warkentin, M. (2002). Issues in mobile e-commerce. Communications of the association for information systems, 8(1), 3.
15. Taniar, D. (Ed.). (2007). Encyclopedia of mobile computing and commerce. IGI Global.
16. Thoi, M. (2016). Exploring merchants' adoption of mobile payments A qualitative study on Swedish merchants' perspectives (Master's thesis).
17. Laudon, K. C., & Traver, C. G. (2016). *E-commerce: business, technology, society*.
18. Nielsen, J. (1994). Heuristic evaluation. In J. Nielsen and R.L. Mack (Eds.), Usability Inspection Methods. John Wiley & Sons, New York, NY
19. Kurosu, M. (2015, August). Usability, quality in use and the model of quality characteristics. In *International Conference on Human-Computer Interaction* (pp. 227-237). Springer, Cham.
20. Olsina, L., Lew, P., Dieser, A., & Rivera, B. (2012). Updating quality models for evaluating new generation web applications. *Journal of Web Engineering*, *11*(3), 209.
21. Parsazadeh, N., Ali, R., Rezaei, M., & Tehrani, S. Z. (2018). The construction and validation of a usability evaluation survey for mobile learning environments. *Studies in Educational Evaluation*, *58*, 97-111.
22. Vidal, L.-A., Marle, F., & Bocquet, J.-C. (2011). Using a Delphi process and the Analytic Hierarchy Process (AHP) to evaluate the complexity of projects. Expert Systems with Applications, 38(5), 5388–5405.
23. Skulmoski, G., Hartman, F., & Krahn, J. (2007). The Delphi method for graduate research. Journal of Information Technology Education: Research, 6(1), 1–21.
24. Athukorala, N. A. (2011). An Empirical Study of the critical success factors for business process re-engineering (BPR) in the employees' provident fund.
25. Hoehle, H., Zhang, X., & Venkatesh, V. (2015). An espoused cultural perspective to understand continued intention to use mobile applications: a four-country study of mobile social media application usability. *European Journal of Information Systems*, *24*(3), 337-359.
26. Hsu, C. W., & Yeh, C. C. (2018). Understanding the critical factors of usability for successful M-commerce adoption. *International Journal of Mobile Communications*, *16*(1), 50-62.
27. Hoehle, H., Aljafari, R., & Venkatesh, V. (2016). Leveraging Microsoft′ s mobile usability guidelines: Conceptualizing and developing scales for mobile application usability. *International Journal of Human-Computer Studies*, *89*, 35-53.
28. Kim, K., Proctor, R. W., & Salvendy, G. (2012). The relation between usability and product success in mobile applications. *Behaviour & Information Technology*, *31*(10), 969-982.
29. Lee, D., Moon, J., Kim, Y. J., & Mun, Y. Y. (2015). Antecedents and consequences of mobile phone usability: Linking simplicity and interactivity to SATS, trust, and brand loyalty. *Information & Management*, *52*(3), 295-304.
30. Moumane, K., Idri, A., & Abran, A. (2016). Usability evaluation of mobile applications using ISO 9241 and ISO 25062 standards. *SpringeYlus*, *5*(1), 548.
31. Balapour, A., & Sabherwal, R. (2017). Usability of Apps and Websites: A Meta-Regression Study.
32. Kellar, S. P., & Kelvin, E. A. (2013). Munro's Statistical MLR Prediction Methods for Health Care Research (6thed.). Philadelphia, PA: Wolters Kluwer Health | Lippincott Williams and Wilkins. Chapter 14.
33. Sarro, F., Petrozziello, A., & Harman, M. (2016, May). Multi-objective software effort estimation. In *2016 IEEE/ACM 38th International Conference on Software Engineering (ICSE)* (pp. 619-630). IEEE.